\documentclass[12pt,preprint,epsf]{aastex}

\def\spose#1{\hbox to 0pt{#1\hss}} 
\def\simlt{\mathrel{\spose{\lower 3pt\hbox{$\mathchar"218$}}      
\raise 2.0pt\hbox{$\mathchar"13C$}}} 
\def\simgt{\mathrel{\spose{\lower 3pt\hbox{$\mathchar"218$}}      
\raise 2.0pt\hbox{$\mathchar"13E$}}} 
\def\eg{{\rm e.g.~}}  
\def\etal{{\rm et~al.~}} 

\lefthead{Mould \& Spitler} 
\righthead{NGC 4594} 
\begin{document}
\title{The stellar population and metallicity distribution of the Sombrero galaxy}

\author{Jeremy Mould} 
\affil{School  of Physics, University of Melbourne, Vic 3010, Australia} 
\authoraddr{E-mail: jmould@unimelb.edu.au} 

\author{Lee Spitler} 
\affil{Centre for Astrophysics \& Supercomputing, Swinburne University,
Hawthorn, Vic 3122, Australia} 
\authoraddr{E-mail: lee.spitler.astro@gmail.com}

\begin{abstract}
Hubble Space Telescope studies of the resolved stellar population of elliptical galaxies have shown that the galaxies form by steady
accretion of gas which is all the while forming stars and evolving chemically to a metallicity distribution that is
as high as solar composition in the most massive objects that have been analyzed, and much lower for low mass
ellipticals. In this paper we study for the first time the stellar content of an early type spiral galaxy, the
massive disk galaxy, the Sombrero, NGC 4594. %The only other %*late type* 
%disk galaxy probed in this way is the Milky Way analog, NGC 891. 
We consider whether the metallicity distribution function (MDF) in the observed field matches that of elliptical galaxies of some luminosity,
and what these data imply for the accretion and enrichment model that can be fitted to the MDF. 
The MDF of NGC 4594 is similar to that of the elliptical galaxy of similar
luminosity, NGC 5128.
The field we are probing is a combination of the galaxy's bulge and halo.
\end{abstract}


\section{Introduction}
Deep imaging of nearby elliptical galaxies has in recent years opened up their 
stellar populations for analysis of the distribution of metallicity. \cite{hh07}
has shown that the metallicity distribution is measurable in a range of galaxies
from the dwarf spheroidals to giant ellipticals such as NGC 5128, and that the 
the metallicity distribution functions (MDF) are sequenced in luminosity,
such that Draco is the most metal poor in mean Z, L* galaxies are intermediate,
 and NGC 5128 the most metal rich. 
In massive galaxies the MDF in metallicity peaks below solar Z and has an extended tail to very low metallicities.  
Lower mass galaxies show a Gaussian MDF at lower metallicities, see \eg \cite{li}.
The MDF constrains models of elliptical
galaxy formation, including monolithic collapse models, hierarchical merging
models, and merger scenarios over cosmic time, see \eg \cite{ikuta}.

Nearby edge-on galaxies present a unique opportunity to learn about the formation
history of the bulge, thick disk, and halo of galaxies. Seth, Dalcanton, \& de Jong (2005)
and Mould (2005) have examined HST imaging of two samples of late type edge-on
galaxies and found thick disks without metallicity gradients and without
asymptotic giant branch stars above the red giant branch tip. These observations
rule out models of slow collapse and of young disk heating as thick disk
formation mechanisms. Models that would be favored include those of Brook
et al (2004) whose chemo-dynamical cosmological simulations identify thick
disk formation with a z $>$ 2 epoch of chaotic merging of multiple peer disks
of mixed gas and stars. 

Bulge dominated disk galaxies can be studied in the edge-on case in the
same way. A superb example is the Sa galaxy NGC 4594. We present such a study
in this paper and compare it with the Milky Way analog, NGC 891, \citep{rej09,iba09}, as well as with its elliptical siblings. 
There is no evidence of a thick disk in the perpendicular
profiles of \cite{bu86}, whose scale height is approximately 400 pc.
So we concentrate on the bulge and halo. The Sombrero galaxy
is the first early type spiral to be so studied, although, in a not-so-edge-on case, we note the work
of \cite{r09} on M31, which is limited to the outer halo. \cite{moh05} has
studied the disk galaxies NGC 55, 247, 253, 300, 3031, 4244, 4258, and 4945.
They find that the MDFs are dominated by metal-rich populations,
with a tail extending toward the metal poor end and that, to first order, the MDFs
are similar to what is predicted by simple, single-component model of chemical
evolution with the effective yields increasing with galaxy luminosity.

\section{Photometry}
The HST archive contains six undithered F606W WFPC2 exposures of a field in the
Sombrero galaxy and six exposures in F814W, amounting to 11,700 seconds in
each filter. These were obtained as part of the medium deep survey of
the original HST Key Projects, GO program 5369. The field is 6.8 effective
radii from the center of the galaxy and is located as depicted in Figure 1.
According to \cite{young} $\sim$90\% of the luminosity of an r to the quarter law
spherical bulge is contained within this radius, and so we can expect a significant
fraction of the stars in this field to be halo stars.

The frames were combined with IRAF and processed with DAOPHOT
\citep{st87}. A sample of the field is shown in Figure 2, allowing
an impression of the level of crowding. Point spread functions were borrowed from the extragalactic
distance scale Key Project \citep{k95}
and fitted to the images with ALLSTAR. The data underwent a second cycle 
with stars found in the residual image. The ALLSTAR files were then combined
with DAOMASTER. The number of stars detected on the four chips was
3647, 10,030, 10,158 and 11,852, where chip 1 is the planetary camera.
Nine bright stars were used to determine aperture corrections
to an accuracy of 0.01 mag. The calibration of \cite{do00} was employed
to transform the color magnitude diagram (CMD) to the VI system.
There are spatial density gradients across the WFPC2 field in the point sources detected of 10--20\%. Both the PC and WF chips are used. The stellar density
increases in the field in the direction towards the center of the galaxy.

The CMD is shown in Figure 3. Stars have been excluded if their color
uncertainties exceed 0.5 mag.
The CMD shows the tip of the red giant branch (TRGB) of NGC 4594. The TRGB
is at I = 26.2 $\pm$ 0.2 mag, corresponding to $(m-M)_0$ = 30.15 $\pm$ 0.2 mag
(see \cite{ms08}). 
%The CMD shows a significant increase in point-like objects at a magnitude of I $\approx$ 26.0.  At the distance of the Sombrero, 
%the TRGB is expected to peak at this magnitude, suggesting we have indeed resolved the brightest magnitude of the TRGB. 
The population is an old one.  

NGC 4594 is at a distance of 29.95 $\pm$ 0.18 mag according to surface
brightness fluctuation measurements by \cite{to01}. This is the distance we adopt. The reddening is
E(V-I) = 0.071 from \cite{sch}. From the distribution
of giant branch colors a magnitude or so fainter than
M$_I$ = --4.0, it is thus possible
to measure the MDF, following \cite{hh07}, who provide template giant
branches for this purpose. The MDF is obtained by interpolation directly
in color-magnitude space and is shown in Figure 4. The isochrone age was 12 Gyrs
and the isochrone spacing was 0.15 dex in [Fe/H].

Eight hundred
artificial stars with V--I = 1.75 mag and 26 $<$ I $<$ 28 were added
to the F606W and F814W frames, and the measurement process was repeated.
The average color error was 0.42 mag with little dependence on magnitude.
The average I magnitude error was 0.44 mag. The color bias due to blending was measured at
--0.02 $\pm$ 0.03, which is negligible. The process was repeated for V--I = 3,
and the color error rose to 0.54 mag. Since the sensitivity of color to metallicity
increases in the red, this increase is not an issue for the MDF. Incompleteness
affects the MDF at the metal rich end. However, this effect is modest: a factor
of two incompleteness in the solar metallicity bin in Figure 4. But a small high
metallicity tail with [Fe/H] $>$ 0 may have been erased by incompleteness. Figure 5 results from photometry of a further 2400 artificial stars and shows
completeness as a function of I magnitude.

\section{Chemical evolution}

Since the time of \cite{pp75} the MDF has been used to constrain the
chemical evolution of galaxies, starting with the Milky Way. With
deviations from the closed box chemical enrichment assumption it is possible
to fit a metal rich distribution like that of the solar neighborhood
with $dA/dS~\rightarrow~\alpha$, where $A$ is the mass of accreted gas,
$S$ is the mass of stars made, and $\alpha$ is the fraction of low mass long-lived
stars formed in a generation of star formation. \cite{h76} showed that
a metal poor MDF, such as that of Galactic halo globular clusters, could be
formed with gas loss from the system, all with standard values of the
yield, $p$, the fraction of heavy elements, $Z$ created in a generation
of star formation and return to the interstellar medium. His equation (7)
can be differentiated to yield:

$$\frac{dS} {dZ} \propto e^{-(\alpha+c)Z/p} \eqno(1)$$

where $c$ is the ratio of gas lost to stars formed, when star formation
is complete. The resulting MDFs for different values of $c$ are shown
in Figure 6.

We also consider a somewhat more general case in which the MDF can be used
to constrain $A(t)$. If $\mu$ is the gas fraction at time t, we have equation
(34) of \cite{pp75} (mass conservation).

$$\mu(t) = 1 - \alpha S(t) + A(t)\eqno(2)$$

If $Z^\prime$ is the Z of the accreted gas, heavy element accounting (equation (6) of \cite{pp75}) gives us:

$$\mu dZ = \alpha p ~dS - (Z-Z^\prime) dA \eqno(3)$$.

Then the MDF can be written as:
$$\frac{dS}{dZ} = \frac{\mu} {\alpha p - (Z-Z^\prime) dA/dS} \eqno(4)$$

So, given the star formation history, S(t), the accretion history, A(t),
and the Schmidt-Kennicutt law, dS/dt $\propto~\mu^n$, 1 $<~n~<$ 2, one
can predict the MDF and compare with observations. An alternative way to
proceed is to use Z(t) as a proxy for time and parameterize the gas depletion
history.

$$\mu = e^{-Z/(px)} \eqno(5)$$
where $x$ is a free parameter controlling the rate of star formation decline.
We can then use the MDF directly to calculate the accretion rate.

$$(Z-Z^\prime) \frac{dA} {dZ} = \alpha p ~\frac{dS} {dZ} - \mu \eqno(6)$$

Figure 7 shows the solution for the Sombrero's MDF. The accreted gas metallicity
was taken to be Z$^\prime$ = 0.0005 and x was put equal to 0.5. Values
of the yield and $\alpha$ were taken from \cite{pp75}.
The interesting results are that there is an early burst of accretion
and that dZ/dt is fairly steady, consistent with the assumption that Z
is a good proxy for time. If dZ/dt is $\sim$0.02/13 Gyr, the duration of the
initial accretion burst is $\sim$0.17 Gyr. Other choices of parameters
yield similar results, although dZ/dt becomes monotonic decreasing,
if the gas depletion history is rapid.

\section{Discussion}

Our measurement of the Sombrero's MDF is, however, severely affected by
photometric errors. As discussed in $\S3$, these range from 0.42 to 0.54 mag
in color, moving from blue to red on the giant branch. This corresponds
to --1, +0.3 dex at [Fe/H] = --1 and $\pm$0.2 dex at [Fe/H] = --0.4.
 To examine this, we ran a simulation with 800 artificial
stars with I = 26.4 and V--I = 1.75. That is a delta function in metallicity
with [Fe/H] = --0.875. The simulated MDF (generated in exactly the same way as for the real data) 
showed a strong low metallicity wing
but fewer solar metallicity stars than Figure 4. We therefore ran
the metallicity delta function through our dA/dZ calculation. The result
is very similar to Figure 7 with a similar early accretion peak.

It is interesting to note that the peak of the MDF in Figure 4 coincides
with the metal-rich peak of the globular cluster MDF of \cite{hs10}.

\cite{hh07} have shown that the MDFs of the elliptical galaxies they have studied (see their figure 15) form a sequence in increasing luminosity and
increasing mean metallicity. With a luminosity M$_B$ = --21.2 and modal
metallicity [Fe/H] = --0.5, NGC 4594 is similar to NGC 5128 with
M$_B$ = --20.6 and modal metallicity --0.45 \citep{RC3}, \citep{hrh09}. 
The bulge dominated Sombrero
galaxy is thus similar chemically to its equivalent elliptical galaxy
and much more metal rich than the Milky Way analog NGC 891 \citep{rej09}
([Fe/H] $\approx$ --1.0).
This suggests that similar processes are occurring to determine the
chemical composition of early type spiral galaxy halos and elliptical galaxies.

Our models of chemical evolution directed at the MDFs of galaxies are
consistent with the idea
that accretion history and gas loss may be key determinants. Low mass galaxies
are unable to retain their gas and display low mean metallicity.
High mass galaxies accrete substantial gas early in their evolution and
process that gas to a high mean metallicity. The mass sequence of \cite{hh07}'s
MDFs from the Draco dwarf, through the Leo ellipticals to NGC 5128 may thus be explained through the
agency of gas loss and accretion.

We have glibly spoken of the field we have observed as characteristic of NGC 4594.
A full study of the stellar population of the Sombrero would search
for chemical gradients in the galaxy, characteristic of chemical enrichment
during collapse and seek distinctions between the bulge, disk, and
globular cluster MDFs. We leave this for further work.

\section{Conclusions}
The MDF of a field 6.8 effective radii from the center of the Sa galaxy NGC 4594
has a modal metallicity of [Fe/H] = --0.5, similar to that of NGC 5128,
an elliptical galaxy of similar luminosity. This is the first early type
disk galaxy to have its MDF measured.

Such a metal rich MDF can be achieved in accreting chemical evolution models
with an initial burst of gas accretion lasting of order 0.17 Gyrs. This is the
time elapsed between redshifts as high as 14 and 10. Alternatively, a loss
of two thirds of the gas from this field during star formation, say, on to 
the disk could have yielded the observed MDF.

Higher signal to noise CMDs of the Sombrero over multiple fields will help to unravel the
star formation and chemical enrichment history of galaxy bulges.

We thank the referee for comments which have improved the paper.
This work is based on data obtained with the NASA/ESA Hubble Space Telescope.
In addition to DAOPHOT, this research has made use of IRAF. IRAF is distributed
by NOAO which is operated by AURA for NSF.

\begin{figure}[h]
\plotone{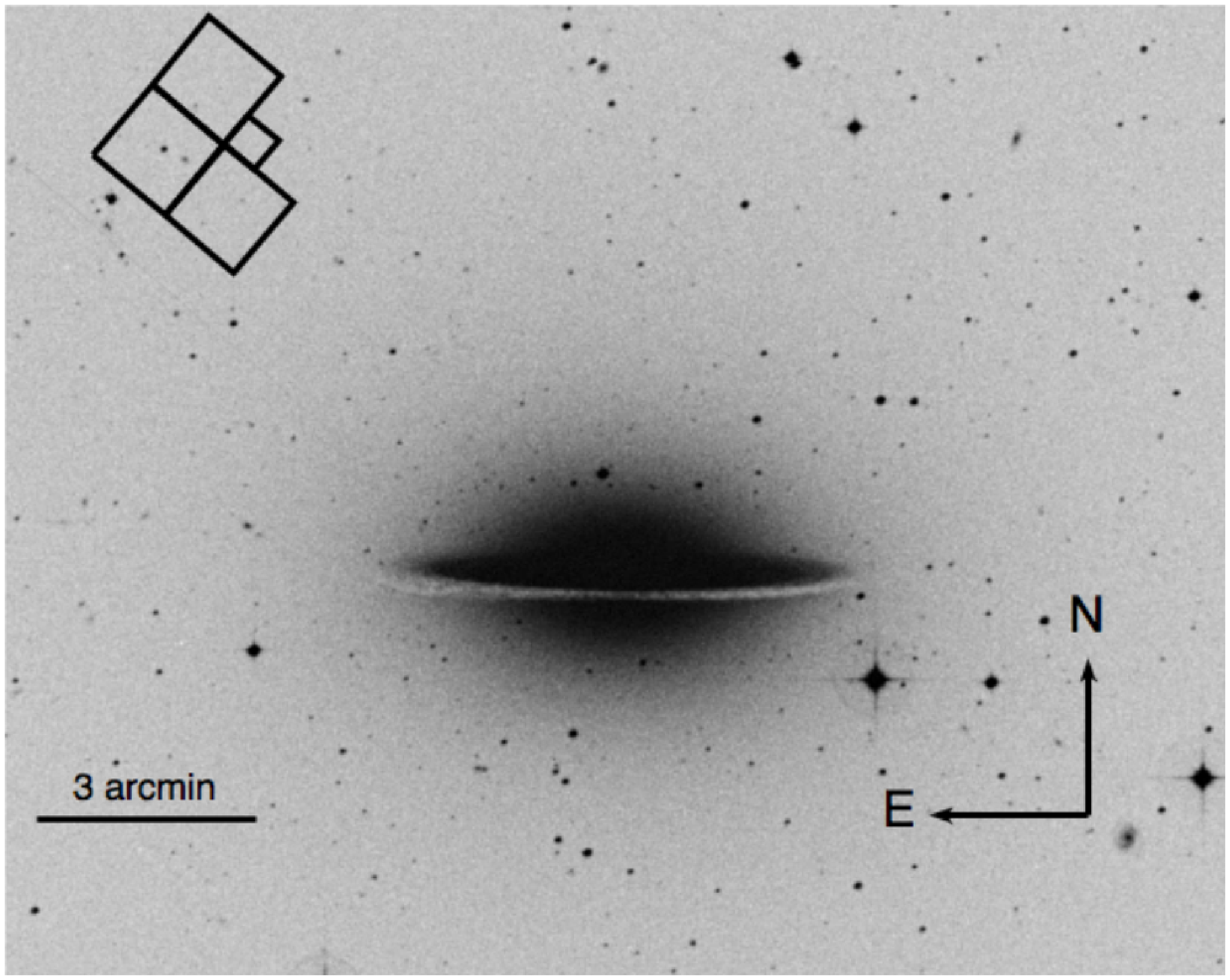}
\figcaption{The field of NGC 4594 with the WFPC2 footprint shown.}
\end{figure}

\begin{figure}[h]
\plotone{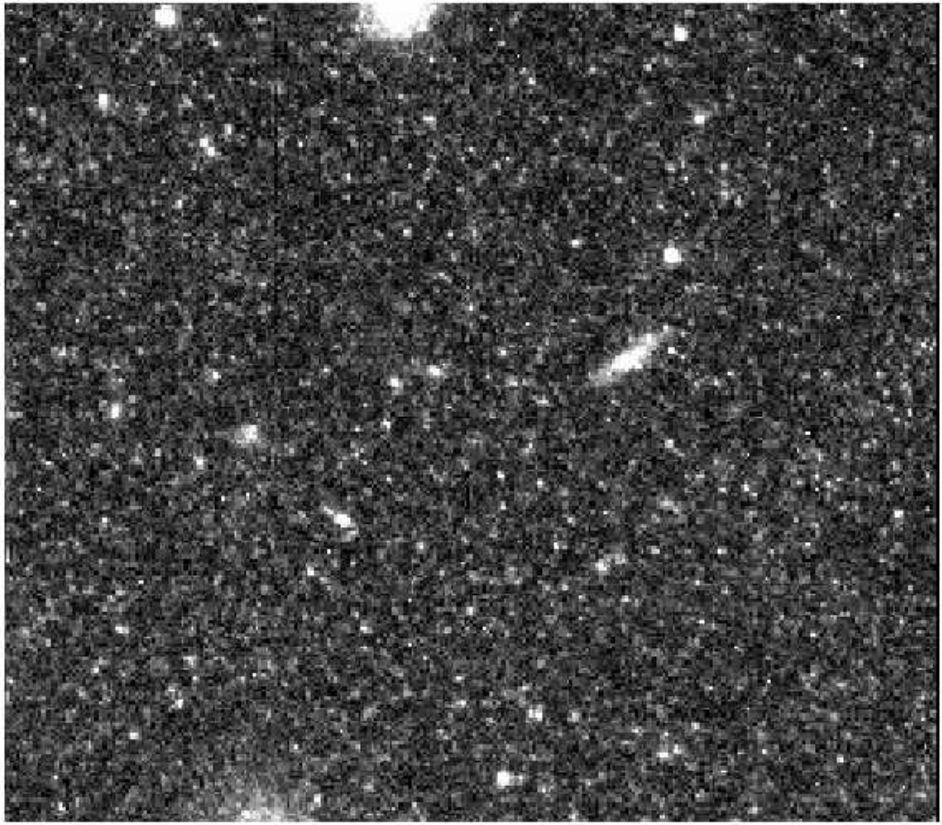}
\figcaption{An enlargement of a 300 $\times$ 300 pixel area in the N4594 field
from the center of chip 2.}
\end{figure}

\begin{figure}[h]
\centering{\includegraphics[angle=-90,width=\textwidth]{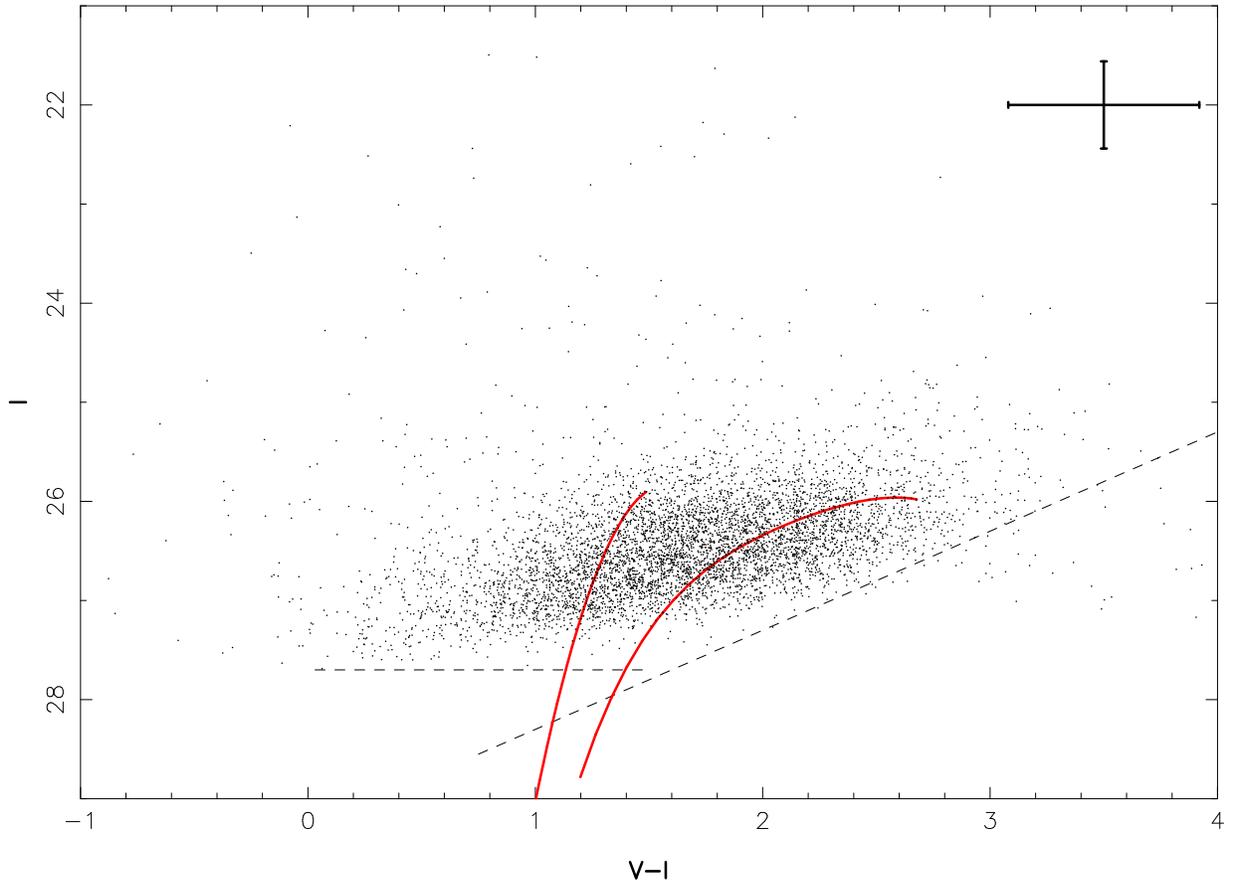}}
\figcaption{Color magnitude diagram of NGC 4594. The solid curves are the
giant branches of M15 and 47 Tuc from \cite{da90} moved to the distance
and reddening of the galaxy. The dashed lines are the 50\% completeness lines
V = 29.3 mag and I = 27.7 mag. M15 has [Fe/H] = --2.17 and 47 Tuc has [Fe/H] = --0.71.}
\end{figure}

\begin{figure}[h]
\centering{\includegraphics[angle=-90,width=\textwidth]{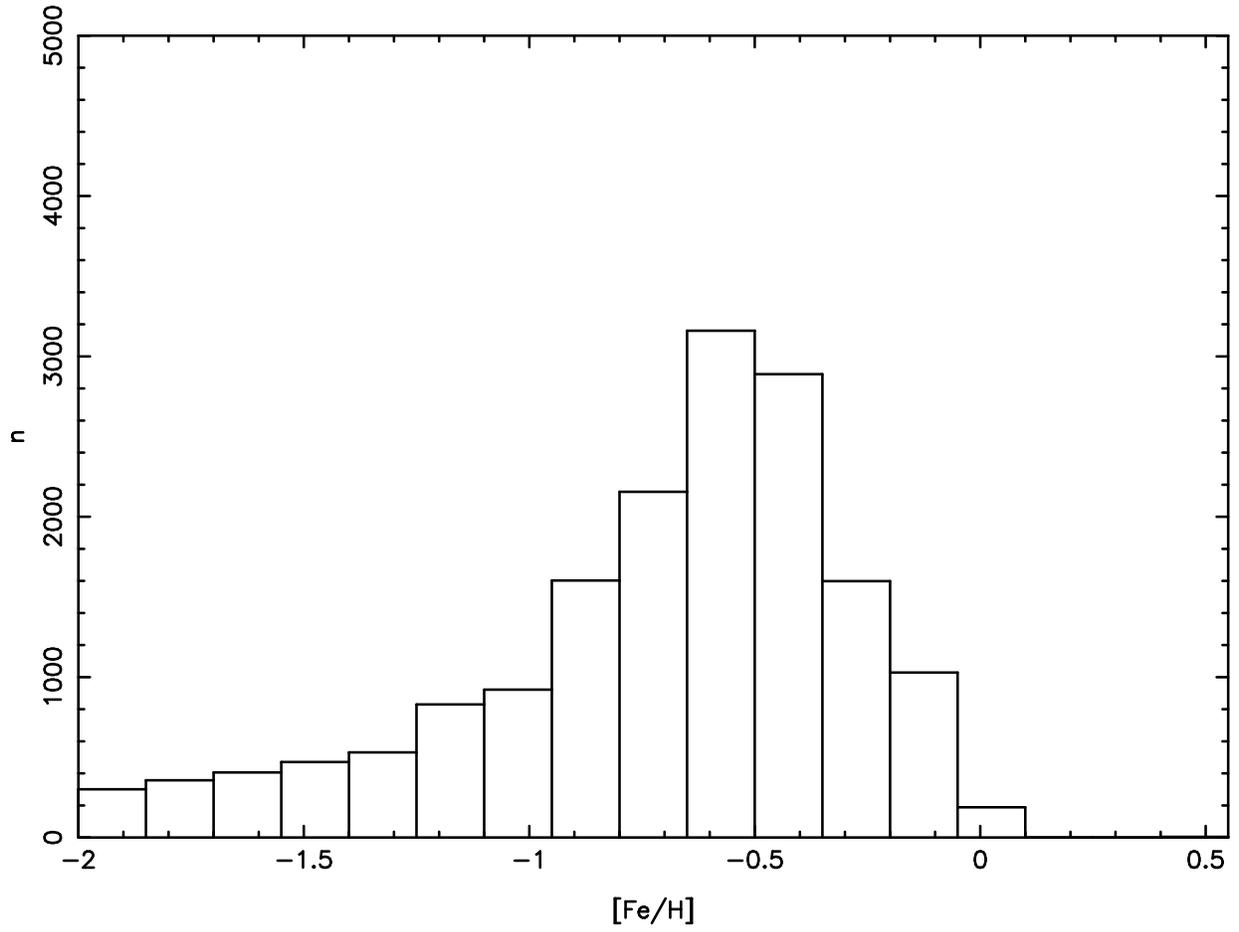}}
\figcaption{Metallicity distribution of the detected red giants in NGC 4594.}
\end{figure}

\begin{figure}[h]
\centering{\includegraphics[angle=-90,width=\textwidth]{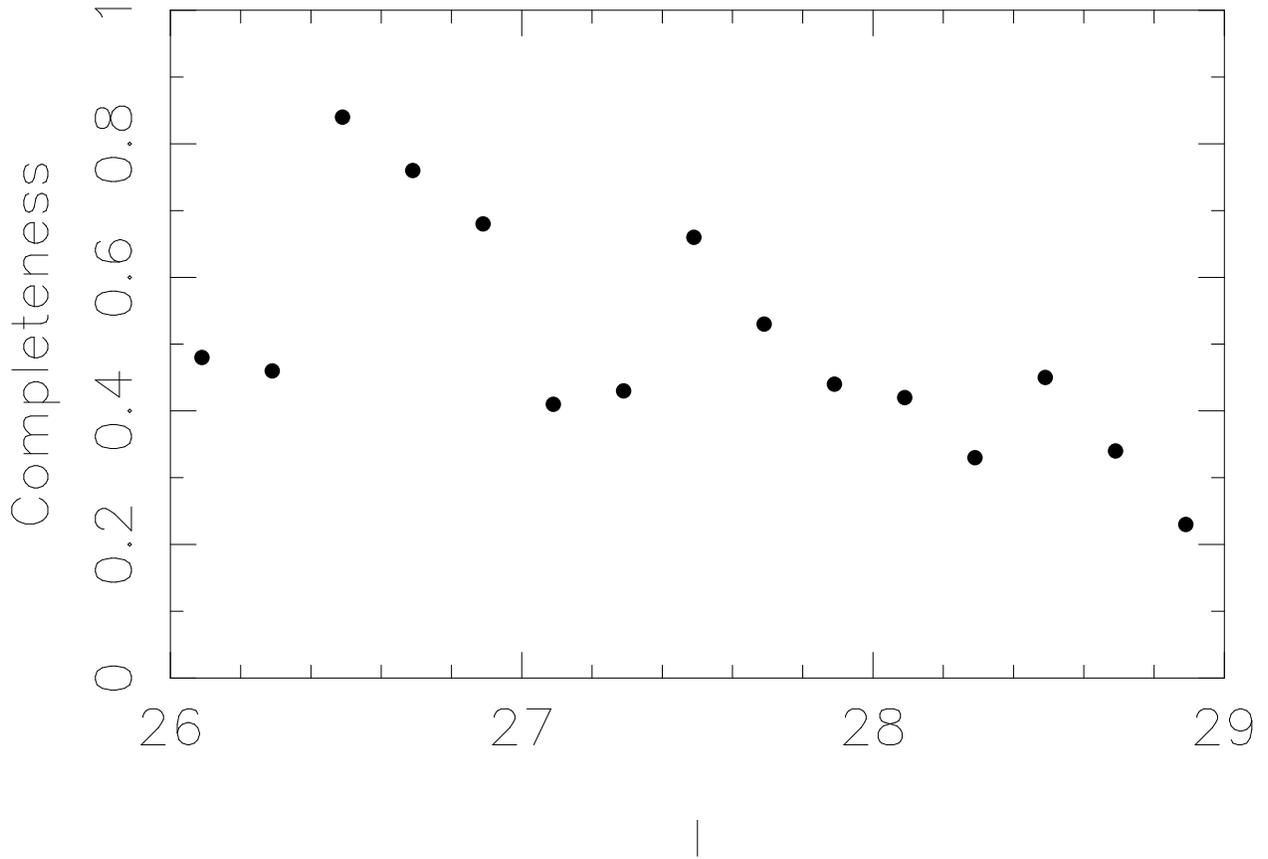}}
\figcaption{Photometric completeness as a function of I magnitude. The 50\%
completeness level shown in Figure 3 is at I = 27.7 $\pm$ 0.3 mag.}
\end{figure}

\begin{figure}[h]
\centering{\includegraphics[angle=-90,width=\textwidth]{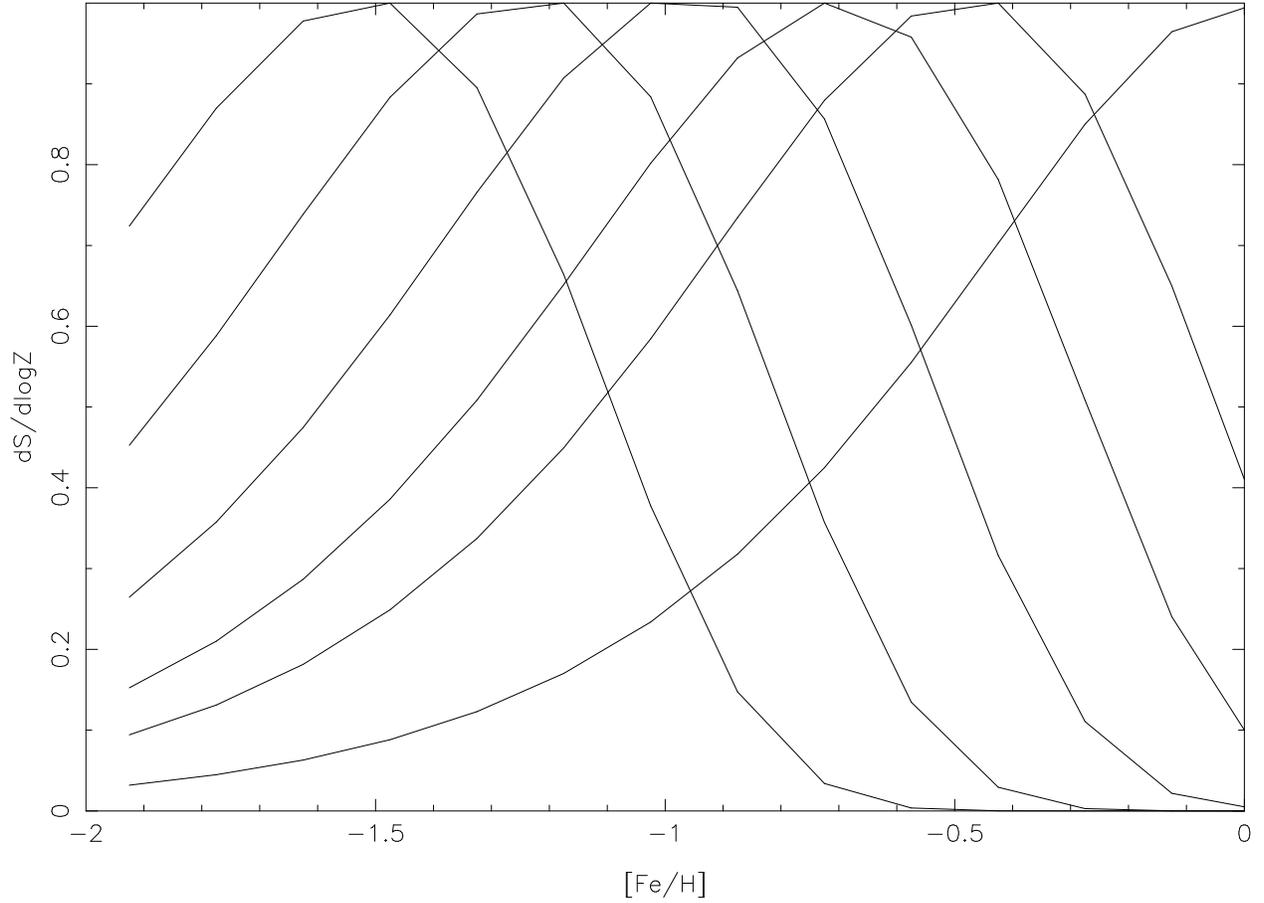}}
\figcaption{Metallicity distribution in the simple model of chemical
evolution with gas loss. The ratio of gas lost to stars formed takes
the values for distributions from left to right of 32, 16, 8, 4 ,2, 0.}
\end{figure}

\begin{figure}[h]
\centering{\includegraphics[angle=-90,width=\textwidth]{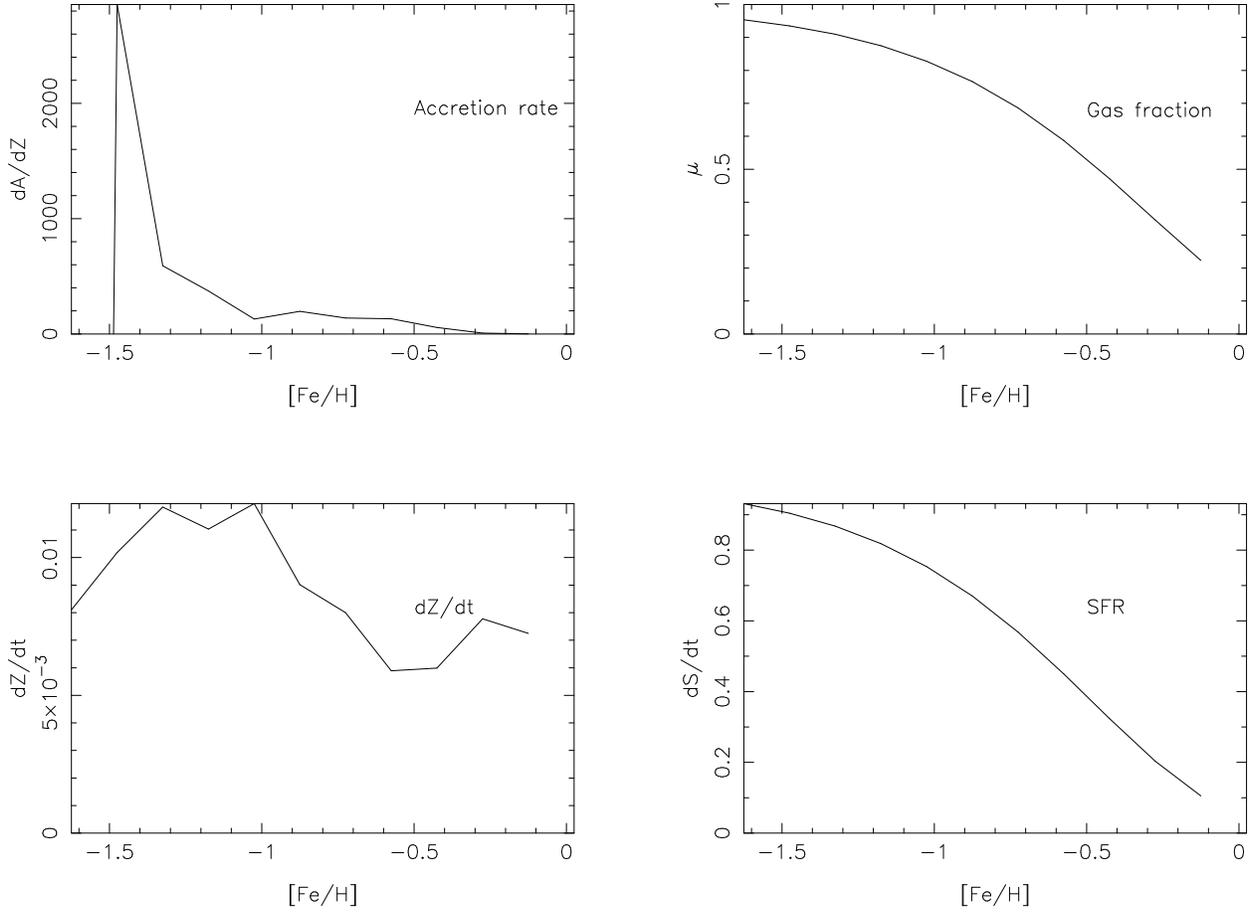}}
\figcaption{The accretion rate as a function of metallicity (top left) in NGC 4594.
The independent variable, [Fe/H] = $\log Z/Z_\odot$, can be considered to be a proxy for time.
The assumed gas depletion rate is depicted in the top right. The chemical
enrichment rate (arbitrary units) is at the bottom left, and the star
formation rate (arbitrary units) at the bottom right.}
\end{figure}

\end{document}